\documentclass[letter]{aa}
\usepackage[T1]{fontenc}
\usepackage[utf8]{inputenc}

\usepackage{newtxtext}  
\usepackage[varg]{newtxmath}  
\usepackage{amsmath}    
\usepackage{amssymb}    
\usepackage{marvosym}   
\usepackage{nicefrac}   
\usepackage{textcomp}
\usepackage{url}
\usepackage{bm}         
\usepackage{mathtools}
\usepackage{microtype}

\usepackage[exponent-product=\cdot
            ,tight-spacing=true
            ,range-phrase={\text{--}}
            ,range-units=single
            ,list-units=single
            ,angle-symbol-over-decimal=true
            ,detect-weight=true
            ,list-pair-separator= {\text{ and }}
            ]{siunitx}
\DeclareSIUnit\au{au}
\DeclareSIUnit\pc{pc}
\DeclareSIUnit\jy{Jy}
\DeclareSIUnit\msun{M\ensuremath{_{\sun}}}
\DeclareSIUnit\lsun{L\ensuremath{_{\sun}}}

\usepackage{graphicx}
    \DeclareGraphicsExtensions{.pdf}
    \setkeys{Gin}{width=\linewidth}
    \setkeys{Gin}{height=\textheight}
    \setkeys{Gin}{keepaspectratio=true}
    \graphicspath{{../../plots/}}
\usepackage{float}
\usepackage{tikz}
\usepackage{tikzscale}

\usepackage{hyperref}
\usepackage{prettyref}  
\newrefformat{fig}{\text{Fig.~\ref{#1}}}
\newrefformat{eq}{\text{Eq.~\ref{#1}}}
\newrefformat{tab}{\text{Table~\ref{#1}}}
\newrefformat{sec}{\text{Sect.~\ref{#1}}}
\newrefformat{enu}{\text{~\ref{#1}}}

\newcommand{\pref}[1]{\prettyref{#1}}

\begin{document}

\title{Polarization reversal of scattered thermal dust emission in protoplanetary disks at (sub-)mm wavelengths}
\titlerunning{Polarization reversal in the (sub-)mm}

\author{R. Brunngräber \and S. Wolf}
\institute{Institut für Theoretische Physik und Astrophysik, Christian-Albrechts-Universität zu Kiel, Leibnizstr. 15, 24118 Kiel, Germany\\\email{rbrunngraeber@astrophysik.uni-kiel.de}}

\date{Received / Accepted}

\abstract{The investigation of polarized light of protoplanetary disks is key for constraining the dust properties, disk morphology and embedded magnetic fields. However, different polarization mechanisms and the diversity of dust grain shapes and compositions lead to ambiguities in the polarization pattern.

The so-called ``self-scattering'' of thermal, re-emitted radiation in the infrared and mm/submm is discussed as a major polarization mechanism. If the net flux of the radiation field is in radial direction, it is commonly assumed that the polarization pattern produced by scattering in a protoplanetary disk shows concentric rings for disks seen in face-on orientation.

We show that a flip of $90^\circ$ of the polarization vectors may occur and mimic the typical pattern of dichroic emission of dust grains aligned by a toroidal magnetic field in disks seen close to face-on. Furthermore, this effect of polarization reversal is a fast changing function of wavelength and grain size, and thus a powerful tool to constrain grain composition and size distribution present in protoplanetary disks. In addition, the effect may also provide unique constraints for the disk inclination, especially if the disk is seen close to face-on.}


\maketitle

%
%
%
\section{Introduction}
    Polarized dust emission of protoplanetary disks and molecular clouds in the (sub-)mm wavelength regime was long believed to originate exclusively from dichroic emission and absorption of non-spherical dust grains that are at least partially aligned by a magnetic field. However, dust growth in protoplanetary disks leads to particles of \si{\um} to \si{\mm} size, and hence to optically thick disks at (sub-)mm wavelengths \citep[e.g.][]{wolf-et-al-2008}. The re-emitted thermal radiation of the dust gets scattered by these large grains and therefore polarized, a process that is commonly refered to as ``self-scattering'' \citep{kataoka-et-al-2015}. This scattering is considered as an origin of the observed polarization of selected protoplanetary disks (e.g. HL\,Tau; \citealt{stephens-et-al-2014,yang-et-al-2016a}). The scattering of thermal radiation produces polarization vectors oriented perpendicular to the dominating direction of the anisotropic local radiation field, i.e. it is expected to show concentric rings for face-on protoplanetary disks in the case of the Rayleigh limit \citep{yang-et-al-2016a,yang-et-al-2016b}. Furthermore, \citet{yang-et-al-2016a,yang-et-al-2017} investigated several properties of polarized radiation resulting from scattering in the (sub-)mm wavelength range for example the near-far-side asymmetry for moderatly inclined disks. These studies were performed considering mostly single grain sizes and analytical temperature distributions, and are limited to special cases concerning dust composition and disk geometry that might not reflect the true nature of protoplanetary disks. The consideration of self-consistently calculated temperatures, grain size distributions, multiple scattering, multi-wavelength effects and the exploration of a broad disk and dust parameter space are only possible with the full extent of Monte-Carlo radiative transfer simulations which have hardly been performed to date \citep[e.g.][]{kataoka-et-al-2016,yang-et-al-2017}. However, a large number of recently published observations of polarized radiation rely on the results of these early, limited studies and try to explain the observed patterns with (sub-)mm scattering \citep{ohashi-et-al-2018,lee-et-al-2018,hull-et-al-2018,bacciotti-et-al-2018}.
    
    In this letter, we present the effect of a polarization reversal by \SI{90}{\degree} which is the natural outcome of the anisotropic behaviour of the scattering function in the case of Mie scattering \citep{daniel-1980,fischer-et-al-1994}. This effect may mimic the polarization pattern of aligned dust grains in a toroidal magnetic field for disks seen close to face-on and thus completely change the interpretation of the observations. Furthermore, considering the full scattering properties of the dust may provide a unique and powerful tool to constrain several important parameters for dust and disk evolution, such as the spatial distribution of grain size and composition.
    
    In \pref{sec:pol_reversal}, the general effect of the polarization reversal is described. Subsequently, the influence of the observing wavelength and grain size (\pref{sec:size_wave}), grain size distribution (\pref{sec:size_distro}), grain composition and porosity (\pref{sec:comp_por}), and disk inclination (\pref{sec:incl}) is investigated. We conclude our findings in \pref{sec:conclusion}.
    
    This study is performed with version 4.02.01 of the versatile, publicly available, Monte-Carlo 3D radiative transfer code \texttt{POLARIS}\footnote{\url{http://www1.astrophysik.uni-kiel.de/~polaris/index.html}} \citep{reissl-et-al-2016}.
    
\section{Reversal of the polarization orientation due to anisotropic scattering}
    \subsection{General description of the effect}
    \label{sec:pol_reversal}
        The polarization state of light scattered by single dust grains depends on wavelength, grain size and shape, chemical composition, and the geometrical substructure. In a smooth, undisturbed protoplanetary disk observed at  submm to mm wavelengths the radiation field is dominated by the radial component, i.e. the flux density of the thermal re-emission is decreasing monotonously from inside out. In the simplest case, a protoplanetary disk is seen face-on and the scattered intensity is dominated by single scattering. In this case, the typical disk flaring results in scattering angles close to $\SI{90}{\degree}$ for a major fraction of the observed radiation. Under these assumptions it is possible to predict the polarization pattern of a protoplanetary disk directly by the analysis of the Müller matrix $\mathcal{S}$ of the dust. In this case, the degree and orientation of linear polarization are defined by
        \begin{equation}
        \label{eq:p_single}
            p = -\frac{\mathcal{S}_{12}}{\mathcal{S}_{11}}\ .
        \end{equation}
        The scattering plane is defined by the radiation source, the point of scattering and the location of the observer. If $p>0$, the polarization vector of the scattered light is perpendicular to the scattering plane while it is parallel for $p<0$. Thus, a \SI{90}{\degree} flip of the polarization vector (so-called polarization reversal) is expected to occur once $p$ changes sign.
        
\begin{figure}
            \includegraphics[width=0.495\linewidth]{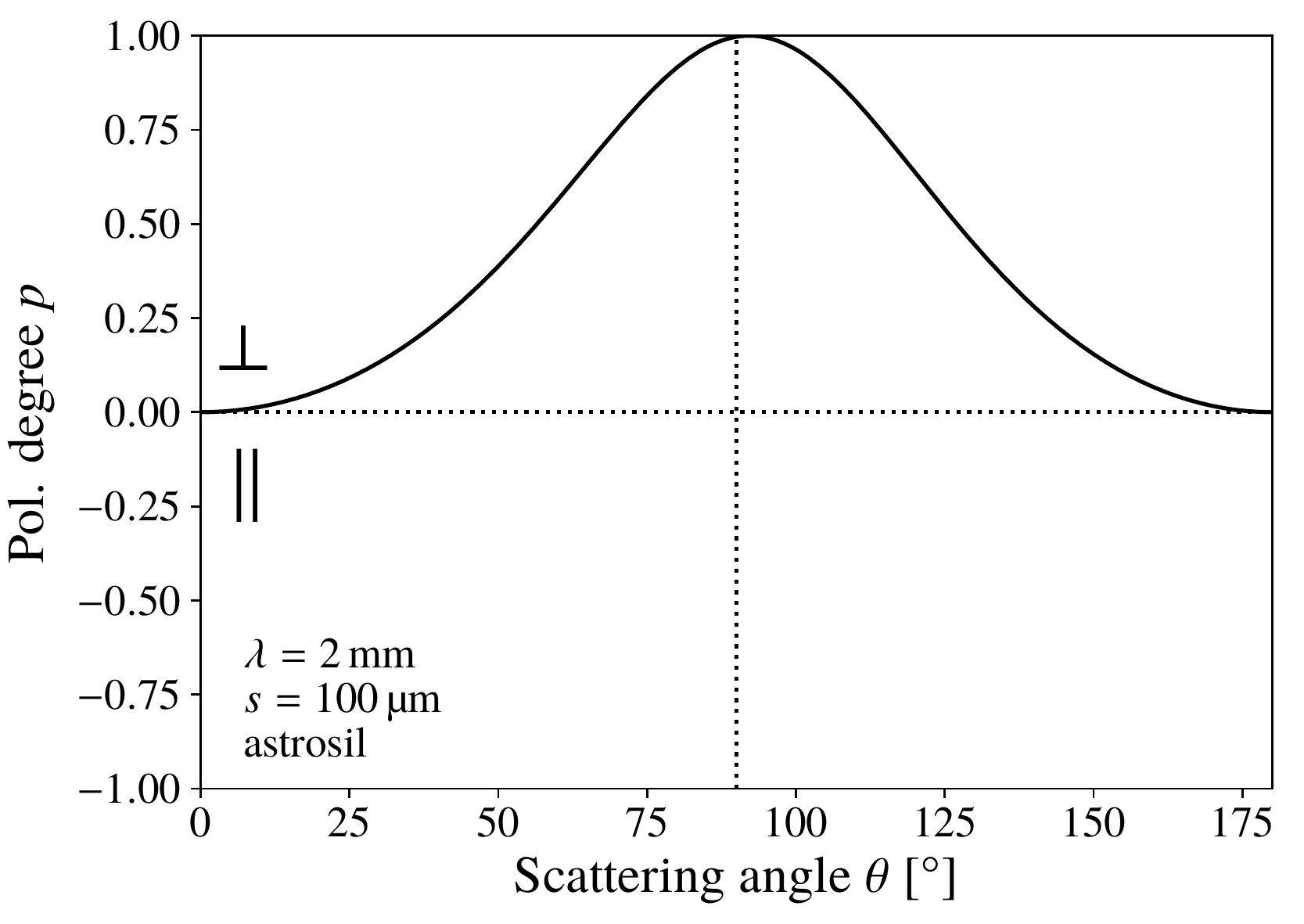}
            \includegraphics[width=0.495\linewidth]{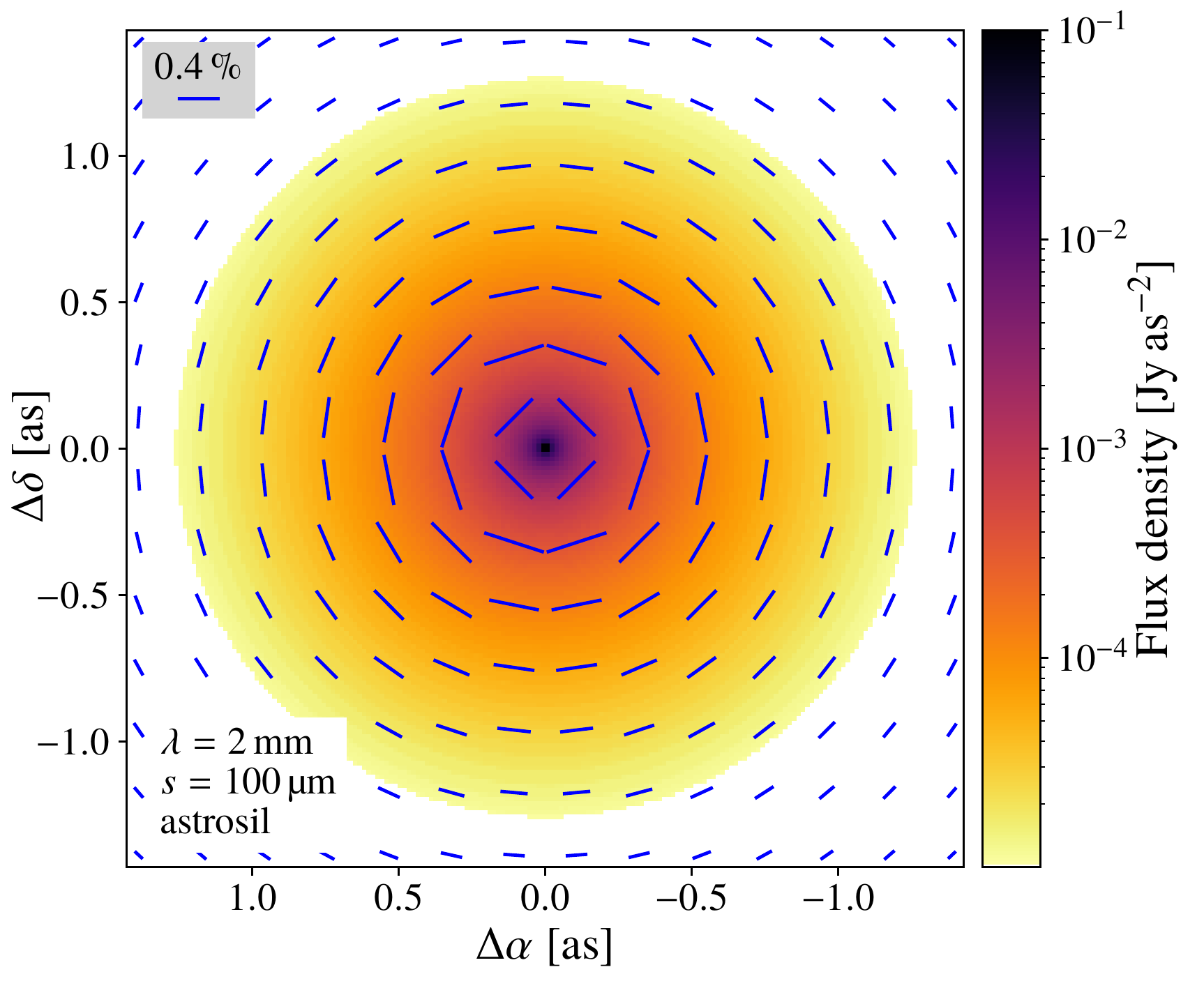}\\
            \includegraphics[width=0.495\linewidth]{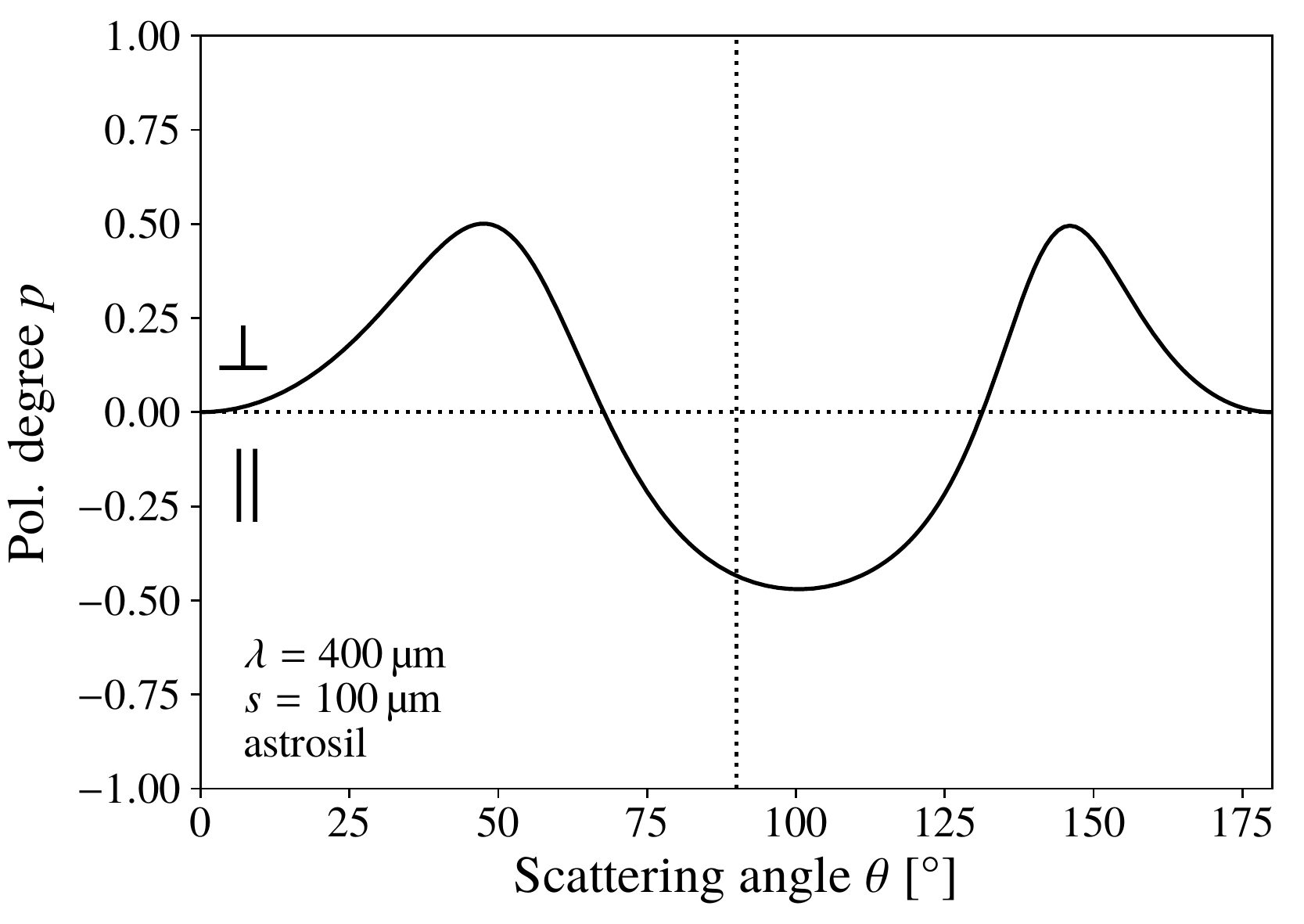}
            \includegraphics[width=0.495\linewidth]{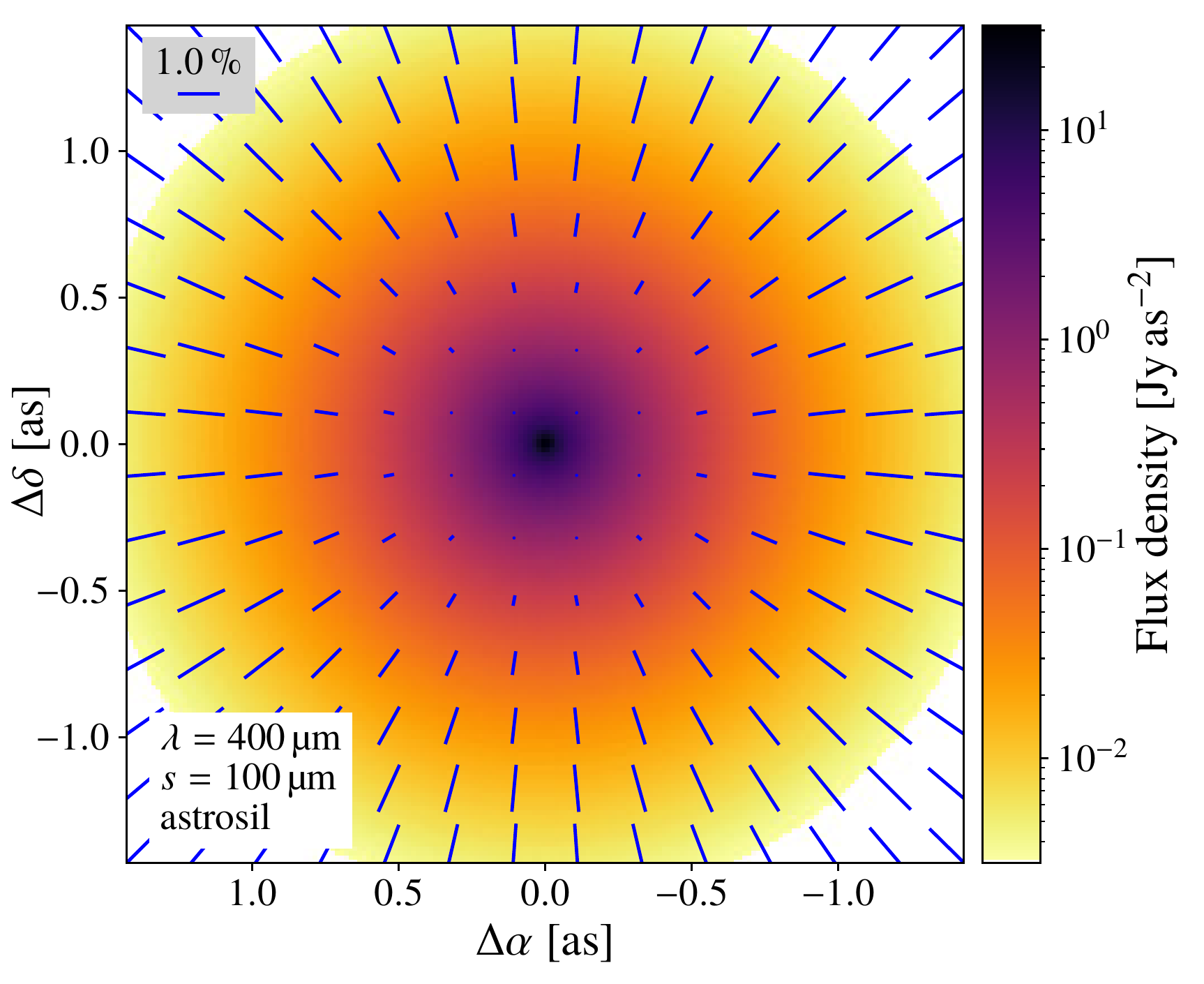}
            \caption{\textit{Left:} Maximum degree of polarization $p = -\nicefrac{\mathcal{S}_{12}}{\mathcal{S}_{11}}$ for single scattering as a function of the scattering angle $\theta$ for astronomical silicate (astrosil). The incident light is unpolarized. The dashed vertical and horizontal lines indicate $\theta=\SI{90}{\degree}$ and $p=0$, respectively. The symbols on the left indicate the orientation of the polarization vector after scattering, i.e. perpendicular ($p>0$) and parallel ($p<0$) to the scattering plane. \textit{Right:} Intensity map of a face-on protoplanetary disk model with superimposed polarization vectors, calculated with the radiative transfer code \texttt{POLARIS} for the case of Rayleigh scattering, i.e. $x = 0.3$ (\textit{top}) and for the case of $x=1.9$ (\textit{bottom}) where a flip of the polarization vectors by \SI{90}{\degree} occurs. These maps include direct re-emission and scattered light. The length of the polarization vectors are scaled to the maximum polarization degree present in each figure.}
            \label{fig:self_sca_general}
        \end{figure}

        The upper row of \pref{fig:self_sca_general} shows the maximum degree of polarization $p$ as a function of the scattering angle $\theta$ for astronomical silicate \citep[refractive indices from][]{draine-lee-1984,laor-draine-1993,weingartner-draine-2001} together with the resulting intensity map and superimposed polarization vectors of a typical density and temperature distribution of a protoplanetary disk (see \pref{sec:disk_setup} for details) in the case of Rayleigh scattering, i.e. where the grain radius $s$ is much smaller than the observing wavelength $\lambda$ and thus the size parameter
        \begin{equation}
            x = \frac{2\pi s}{\lambda} \ll 1\ .
        \end{equation}
        As expected for Rayleigh scattering, the polarization vectors are aligned in concentric rings and are thus oriented perpendicular to the scattering plane. This is the standard case discussed in the literature. \citet{kataoka-et-al-2015} showed that polarization vectors may also have radially outward directed orientations for the Rayleigh regime if the local radiation field is dominated by its azimuthal component (see their Fig.\,7). We find that this case does not occur with the density distribution used here as the resulting intensity distribution has no steep radial gradients except for the regions close the inner rim of the disk which is not spatially resolved in this study. This finding is also confirmed by \citet[their Fig.\,2]{yang-et-al-2016a} who have found polarization vectors which are exclusively aligned in concentric rings for face-on disks.
        
        However, for an increased size parameter, i.e. in the Mie regime, the scattering function is significantly more complex and the resulting polarization degree $p$ is no longer symmetric with respect to forward vs. backward scattering. For size parameters $x\gtrsim1$, $p$ shows a wavy pattern with multiple minima and maxima for different scattering angles. This results not only in different polarization degrees for the different viewing angles, but the quantity $p$ may even become negative which results in a polarization reversal. This effect is illustrated in the lower row of \pref{fig:self_sca_general} for $x=1.9$. While the polarization reversal was found in previous studies \citep[e.g][]{daniel-1980,fischer-et-al-1994,kirchschlager-wolf-2014}, it was not discussed in detail in the terms of (sub-)mm scattering except for a short paragraph in \citet{yang-et-al-2016a}.

        As the Müller matrix elements depend on various quantities, such as grain size, wavelength, chemical composition and grain shape, we emphasize that the effect of polarization reversal may be a common feature and may not be neglected during the analysis of polarization observations of protoplanetary disks. In the following sections, we will briefly discuss the occurrence of polarization reversal as a function of selected parameters.
        
    \subsection{Grain size and observing wavelength}
    \label{sec:size_wave}
\begin{figure}
            \includegraphics[width=0.495\hsize]{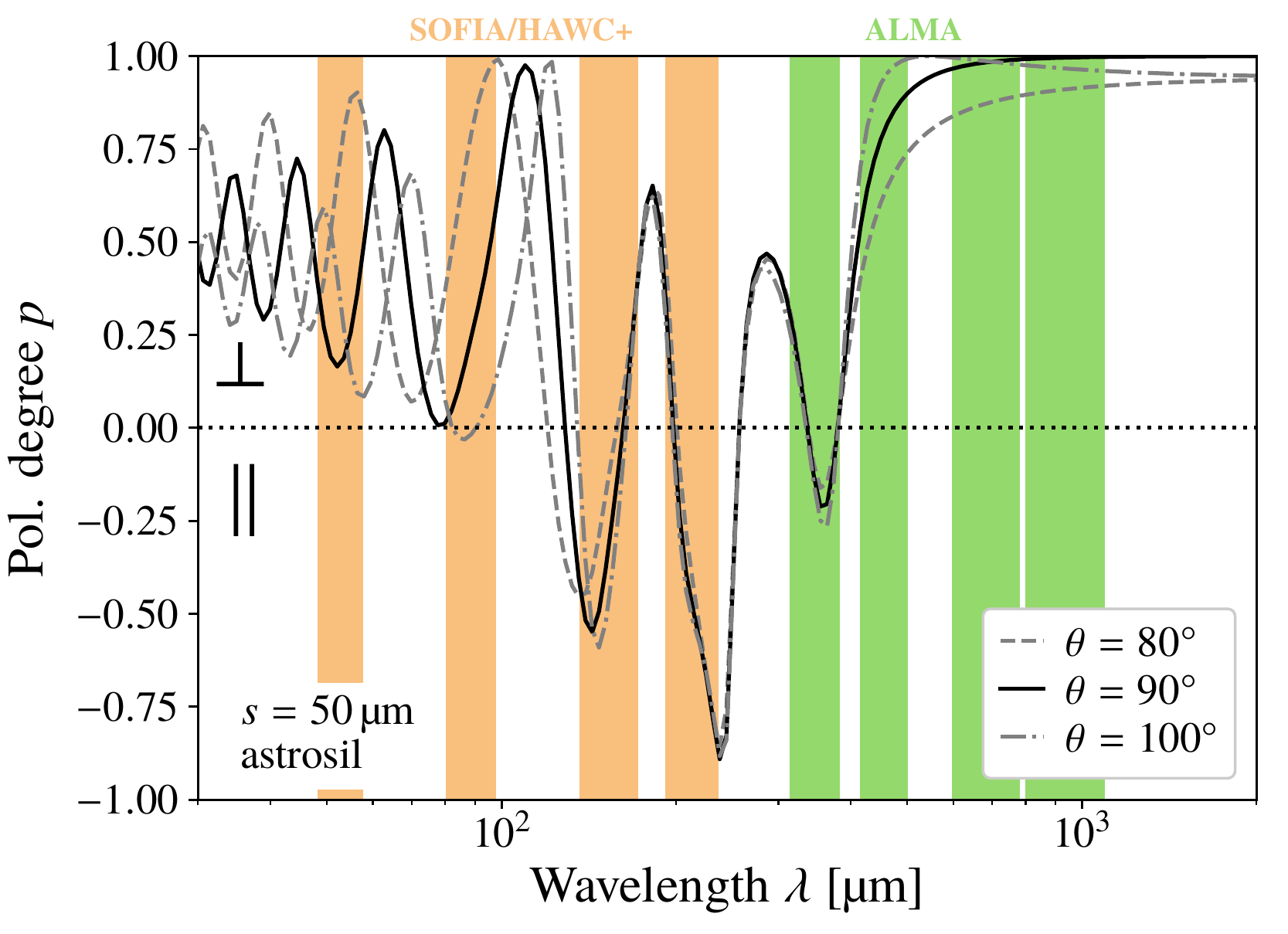}
            \includegraphics[width=0.495\hsize]{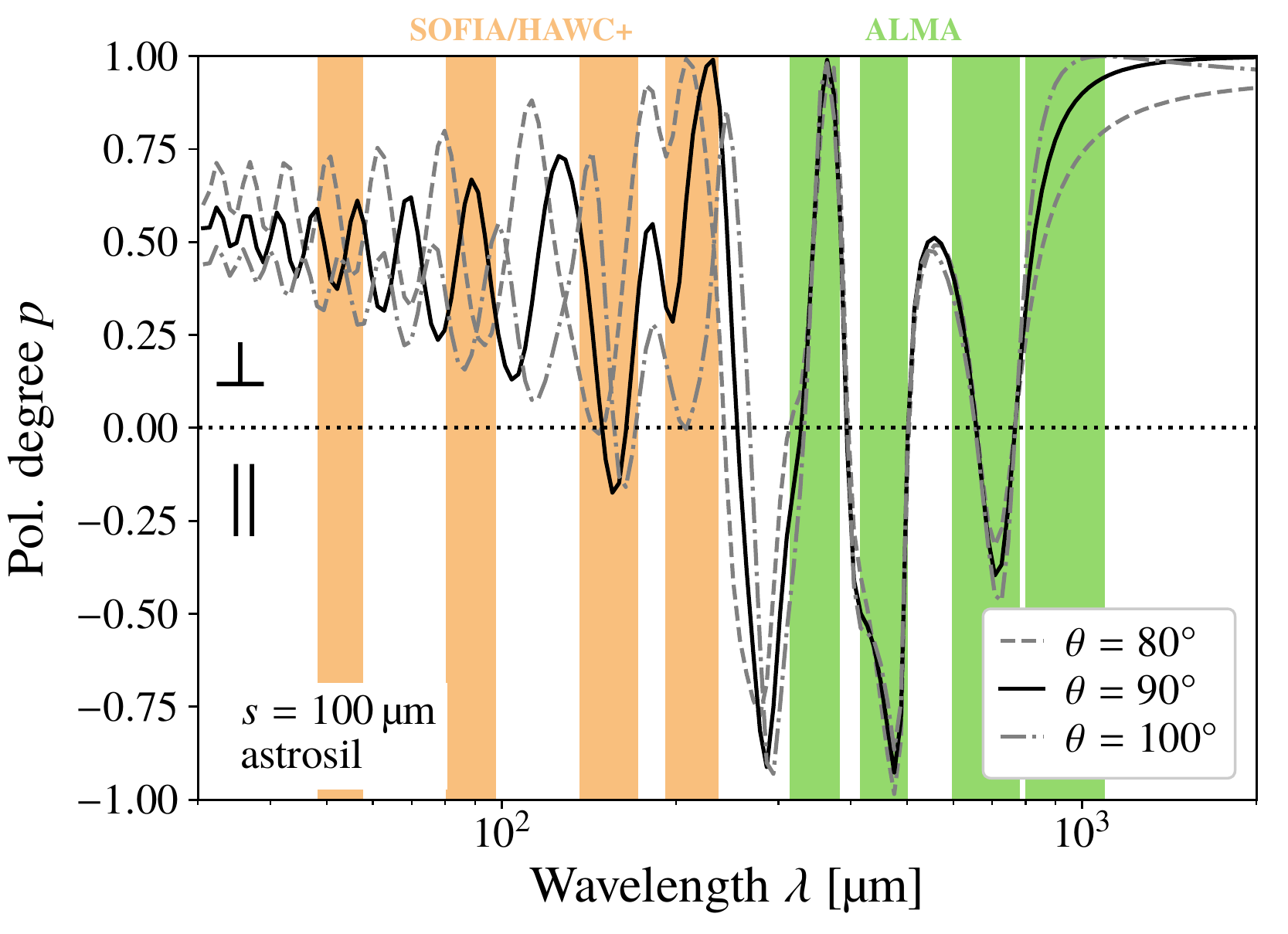}
            \caption{\textit{Left}: Maximum degree of polarization $p$ for silicate grains of size $s=\SI{50}{\um}$ as a function of wavelength for different scattering angles $\theta$ around \SI{90}{\degree}. The orange and green rectangles indicate the bands \textit{A}, \textit{C}, \textit{D}, and \textit{E} of SOFIA/HAWC+, and \textit{10}, \textit{9}, \textit{8}, and \textit{7} of ALMA, respectively. \textit{Right}: Same as left but for grain size $s=\SI{100}{\um}$.}
            \label{fig:pol_frac_90}
        \end{figure}

        Due to the wavelength dependency of the optical properties the polarization degree $p$ is a function of grain size $s$ and wavelength $\lambda$ individually, and of the size parameter $x$. We present the polarization degree for selected grain sizes and wavelengths to emphasize the strong and fast variation of $\nicefrac{\mathcal{S}_{12}}{\mathcal{S}_{11}}$ with variation of these parameters. The polarization fraction for scattering angles of \SIlist{80;90;100}{\degree} is shown for $s=\SIlist{50;100}{\um}$ as a function of wavelength in \pref{fig:pol_frac_90}. In the wavelength range targeted by the far-infrared and mm/submm polarimeters SOFIA/HAWC+ and ALMA between \SI{100}{\um} and \SI{1}{\mm}, the sign of $\nicefrac{\mathcal{S}_{12}}{\mathcal{S}_{11}}$ changes several times. Although SOFIA/HAWC+ operates at the necessary wavelength regime to trace the polarization reversal for \SI{100}{\um} grains, it does not, unlike ALMA, spatially resolve protoplanetary disks.

    \subsection{Grain size distribution}
    \label{sec:size_distro}
        For a grain size distribution the resulting signed polarization degree for single scattering $p$ can no longer be calculated by \pref{eq:p_single}. If multiple grain sizes are present in the disk, the scattering properties must be averaged and weighted by the size distribution $n(s)$ and the different scattering cross sections $C_{\text{sca}}(s)$ of the grains. The signed polarization degree $p$ is thus given by
        \begin{equation}
        \label{eq:p_distribution}
            p = - \frac{\left\langle\mathcal{S}_{12}\right\rangle}{\left\langle\mathcal{S}_{11}\right\rangle}\ .
        \end{equation}
        The averaging process over all grain sizes for a given function $f(s)$ is defined as 
        \begin{equation}
            \left\langle f(s) \right\rangle = \frac{\int_{s_{\text{min}}}^{s_{\text{max}}}{f(s)\cdot n(s)\,C_{\text{sca}}(s)\,\text{d}s}}{\int_{s_{\text{min}}}^{s_{\text{max}}}{n(s)\,C_{\text{sca}}(s)\,\text{d}s}} \ .
        \end{equation}

\begin{figure}
            \includegraphics[width=0.495\hsize]{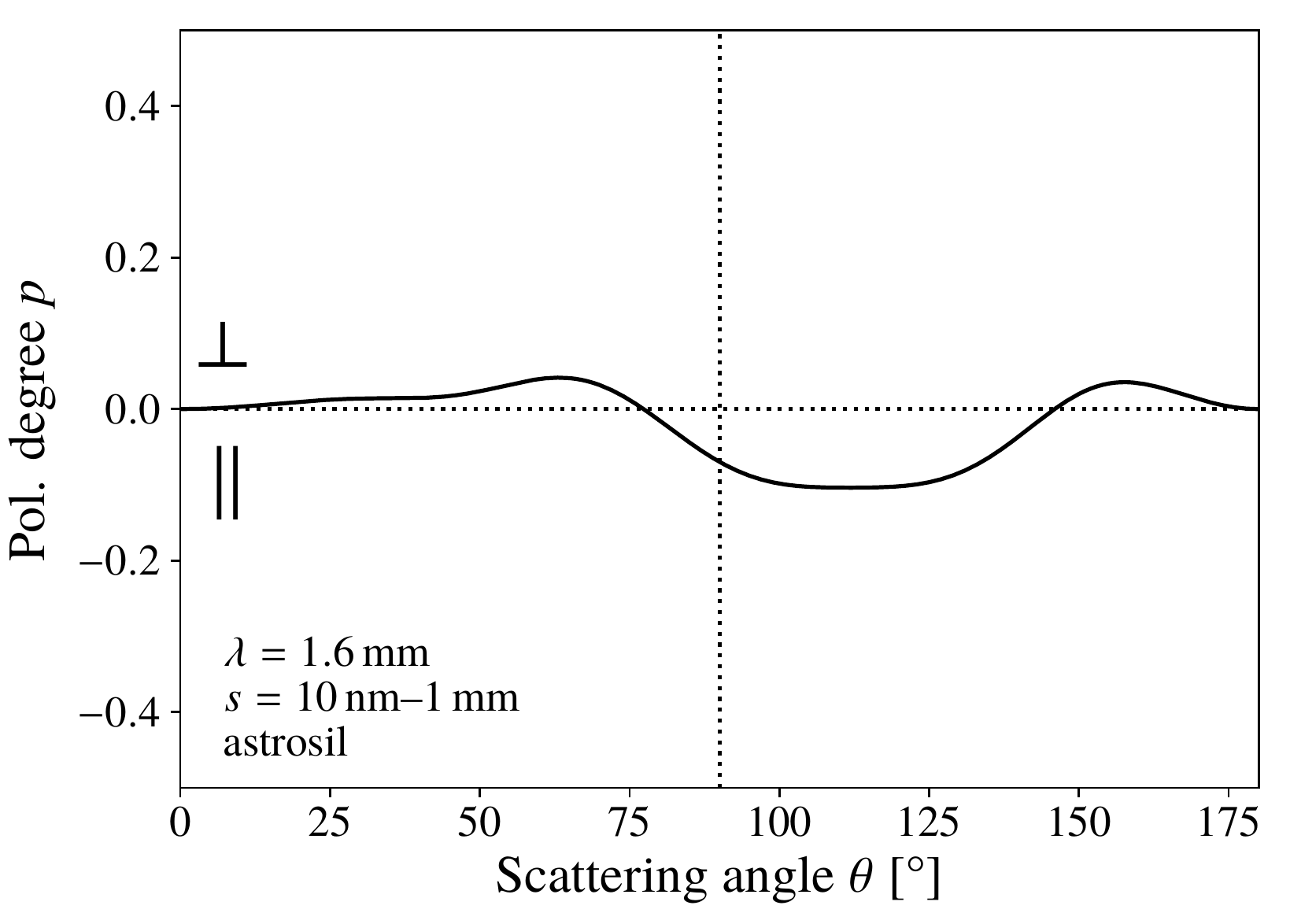}
            \includegraphics[width=0.495\hsize]{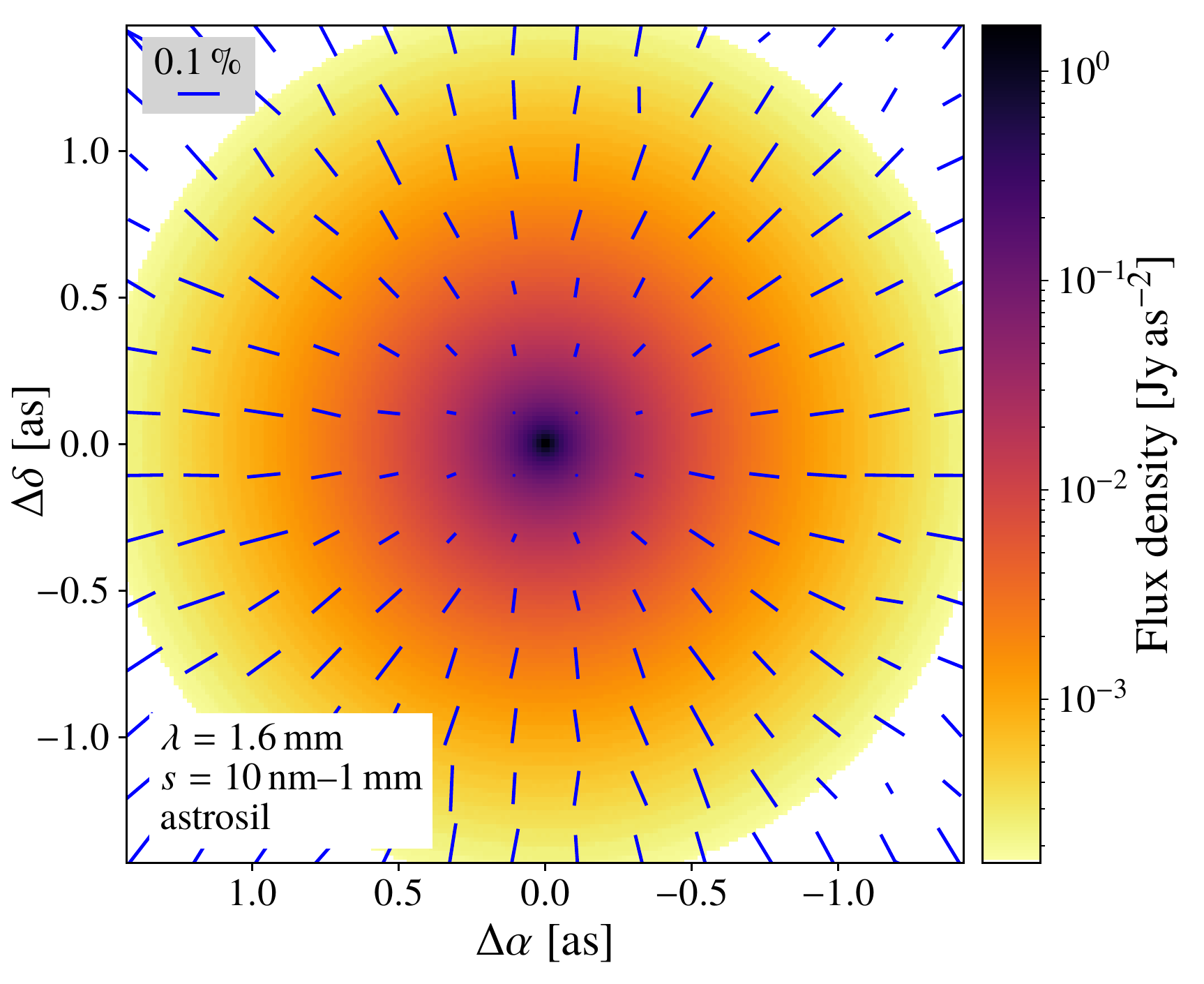}
            \caption{\textit{Left}: Maximum degree of polarization $p$ (cf. \pref{eq:p_distribution}) for silicate grains with sizes between \SI{10}{\nano\m} and \SI{1}{\mm} and a size distribution of $n(s) \propto s^{-2.5}\text{d}s$ as a function of scattering angle $\theta$ at a wavelength of $\lambda=\SI{1.6}{\mm}$. \textit{Right}: Corresponding intensity map of a face-on disk with superimposed polarization vectors. The length of the polarization vectors are scaled to the maximum polarization degree present in the figure.}
            \label{fig:pol_frac_distro}
        \end{figure}

        Considering a grain size distribution rather than single grain sizes will reduce the previously discussed effect of the strongly alternating signed polarization degree $p$ due to the averaging of the Müller matrix elements. The polarization reversal corresponds to a certain size parameter range, thus, observations at different wavelengths must be performed for different grain sizes. Commonly used size distributions for the upper layers of protoplanetary disks span grain radii from \si{\nano\m} scale to hundreds of \si{\nano\m} or at most some \si{\um} \citep{mathis-et-al-1977,glauser-et-al-2008,brunngraeber-et-al-2016,keppler-et-al-2018,casassus-et-al-2018}. For these grain size distributions, the scattering lies in the Rayleigh regime for wavelengths $\lambda \gtrapprox \SI{10}{\um}$. However, with high-resolution polarimeters working in the visible and near-infrared wavelength region such as VLT/SPHERE, the polarization reversal might be detectable for scattered stellar light (see \pref{fig:pol_frac_distro_mixture}). Including even larger grains may still result in a polarization reversal for narrow wavelength ranges in the mm regime. In \pref{fig:pol_frac_distro}, the polarized fraction $p$ and the intensity map of a face-on disk at $\lambda = \SI{1.6}{\mm}$ is shown for grains ranging from \SI{10}{\nano\m} to \SI{1}{\mm} with a power law grain size distribution $n(s) \propto s^{-2.5}\text{d}s$.

    \subsection{Grain composition and porosity}
    \label{sec:comp_por}
\begin{figure}
            \includegraphics[width=0.495\linewidth]{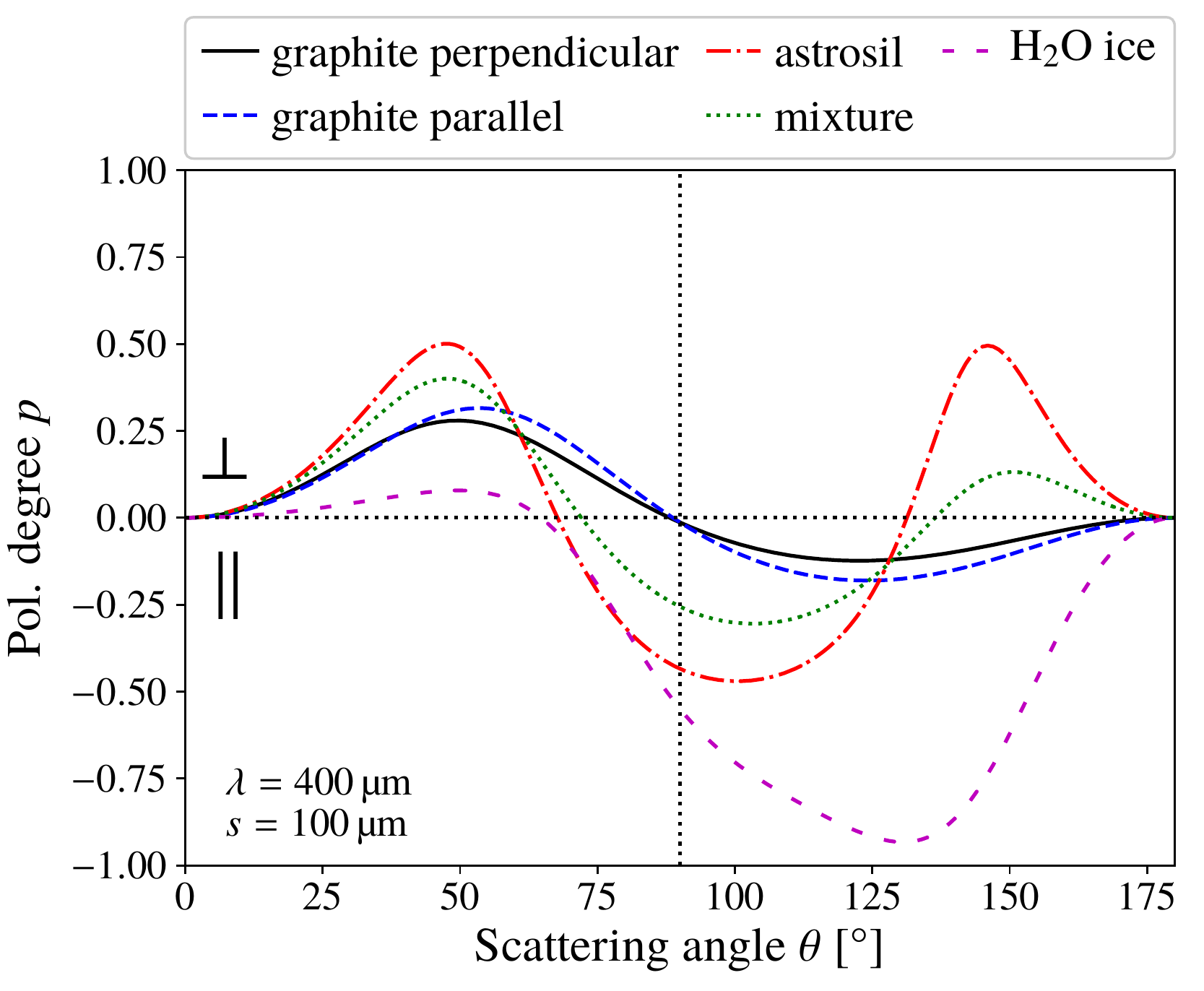}
            \includegraphics[width=0.495\linewidth]{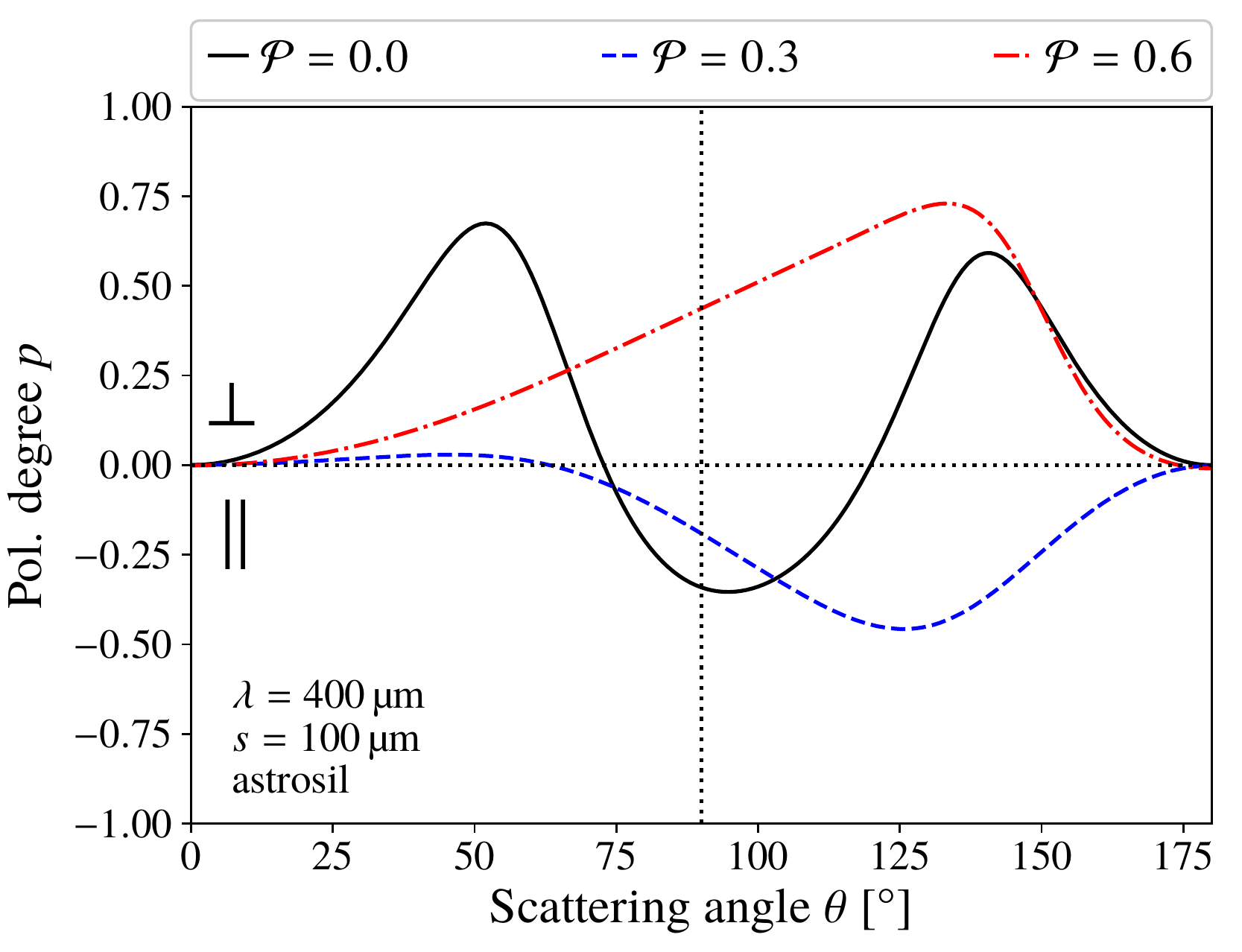}
            \caption{Maximum degree of polarization $p$ of different grain compositions (\textit{left}) and for silicate grains with different porosities (\textit{right}).}
            \label{fig:self_sca_comp}
        \end{figure}

        The most abundant dust grain materials present in the ISM and in protoplanetary disks are silicates and graphite, possibly covered with icy mantles of H$_{\text{2}}$O and other volatile species. Up to this point, all calculations were done assuming pure astronomical silicate. In this section, we will additionally consider pure graphite ($\frac{1}{3}$ parallel and $\frac{2}{3}$ perpendicular, see \citealt{draine-malhotra-1993}; refractive indices from \citealt{draine-lee-1984,laor-draine-1993}), a mixture of silicates (\SI{62.5}{\percent}) and graphite (\SI{37.5}{\percent}) which is a very common assumption for the dust composition in protoplanetary disks \citep{draine-lee-1984,williams-cieza-2011,varga-et-al-2018,brunngraeber-wolf-2018}, and pure water ice grains \citep[refractive indices from][]{warren-brandt-2008}. For pure graphite grains we find that the polarization reversal stops at smaller wavelengths compared to silicates, i.e. the Rayleigh regime extends to slightly larger size parameters $x$. For grain radii of \SI{100}{\um}, the largest wavelengths where the polarization reversal can be observed is about \SI{450}{\um} for graphite and \SI{650}{\um} for silicates. Furthermore, the overall course of the polarization fraction $p$ in the case of graphite is smoother, with shallower gradients. The polarization fraction of a mixture of silicates and graphites shows an intermediate behaviour and changes less rapidly with wavelength and scattering angle if compared to the case of pure silicate, see left panel of \pref{fig:self_sca_comp}.
        
        Another important parameter determining the optical properties of dust grains in protoplanetary disks is porosity, resulting from various grain growth mechanisms \citep[e.g.][]{blum-et-al-2000,ormel-et-al-2008}. We consider spherical, porous silicate dust grains similar to those used by \citet{kirchschlager-wolf-2013,kirchschlager-wolf-2014}. The optical properties were calculated with \texttt{DDSCAT}\footnote{\url{http://ddscat.wikidot.com/downloads}} \citep{draine-flatau-1994}. The polarization fraction $p$ for grains with three different porosities can be seen in the right panel of \pref{fig:self_sca_comp}. Although material and size remain unchanged, vacuum inclusions inside the grain result in a very different scattering behaviour for both different scattering angles and wavelengths.

    \subsection{Disk inclination}
    \label{sec:incl}
\begin{figure}
            \includegraphics[width=0.99\linewidth]{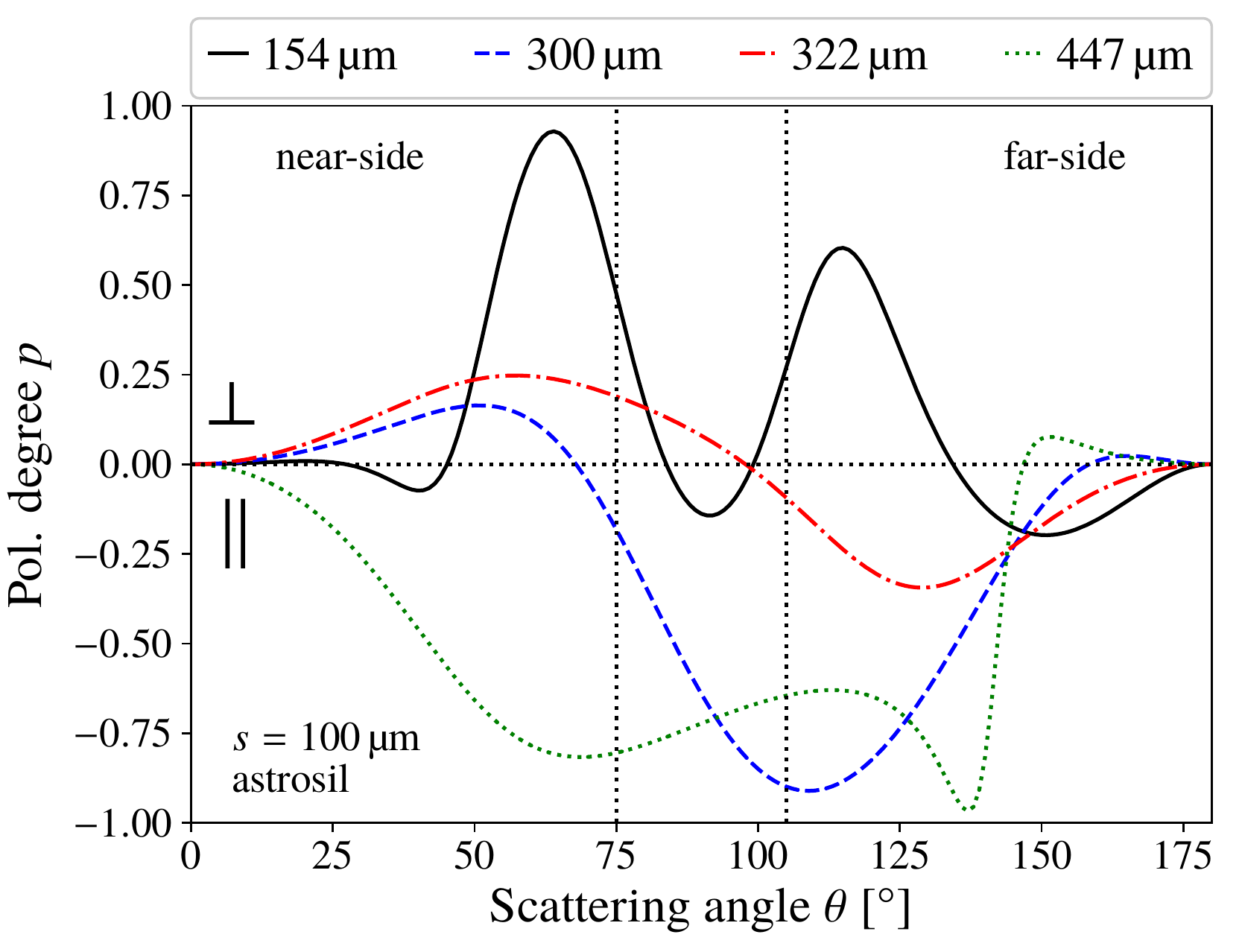}\\
            \includegraphics[width=0.495\linewidth]{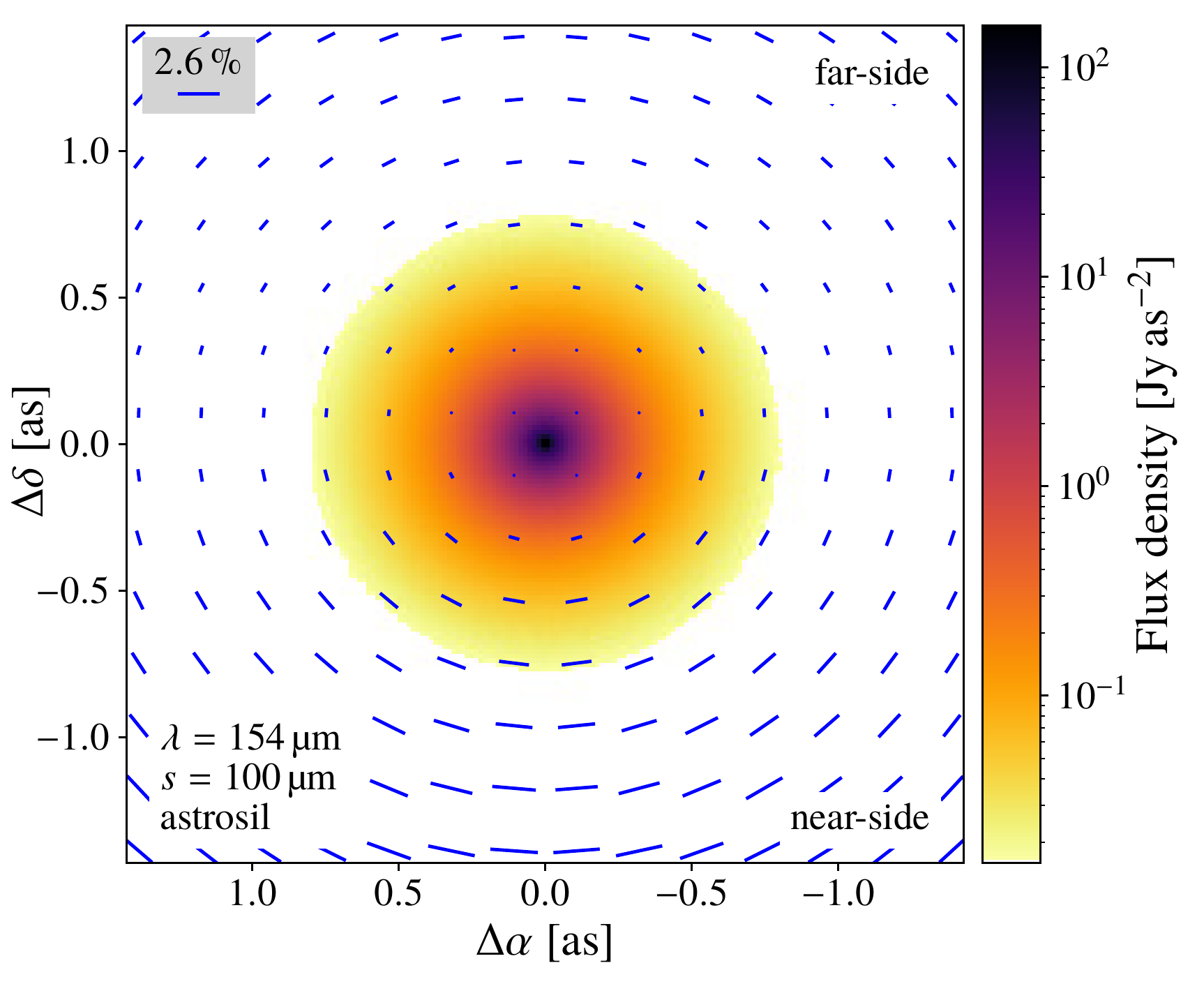}
            \includegraphics[width=0.495\linewidth]{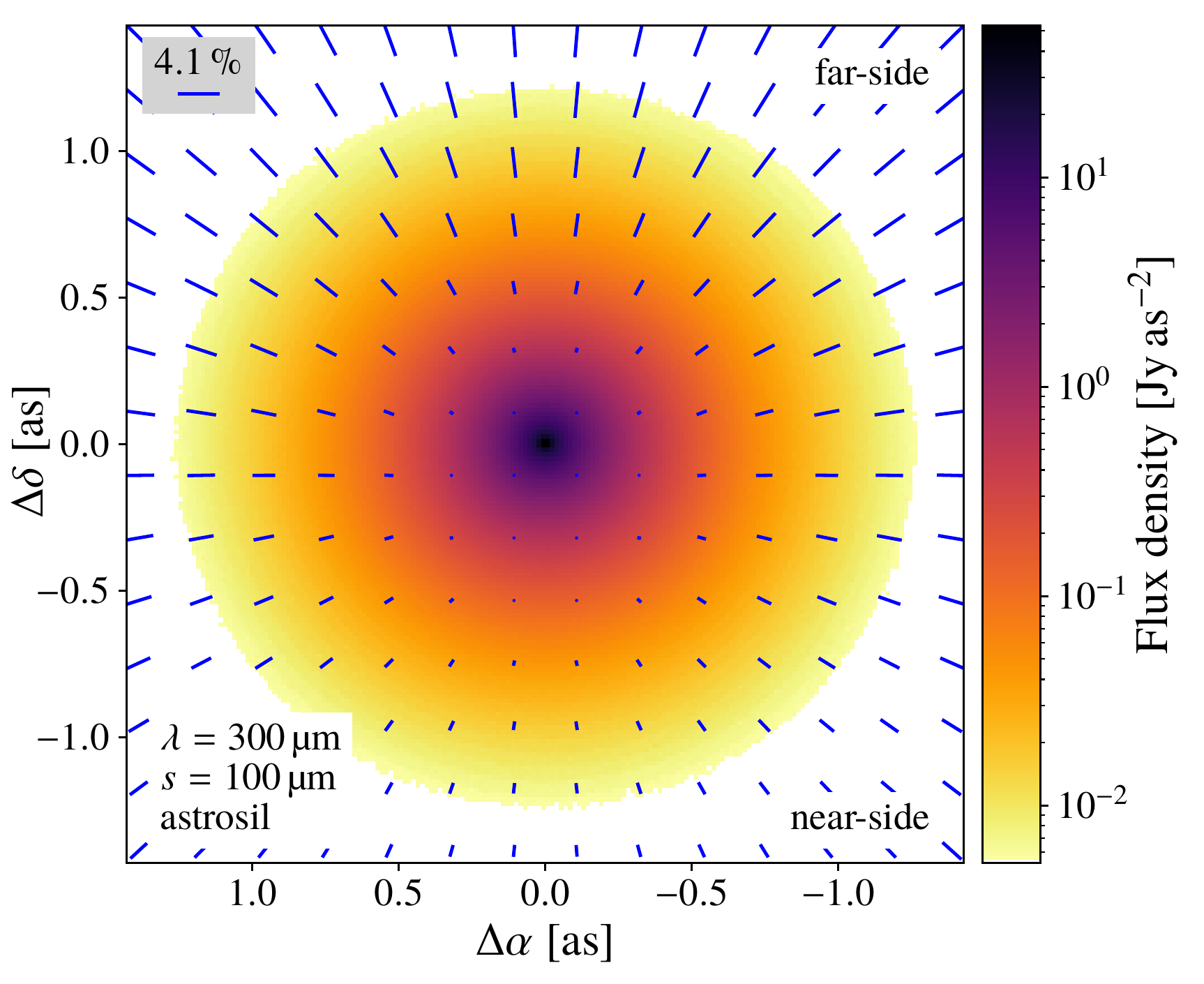}\\
            \includegraphics[width=0.495\linewidth]{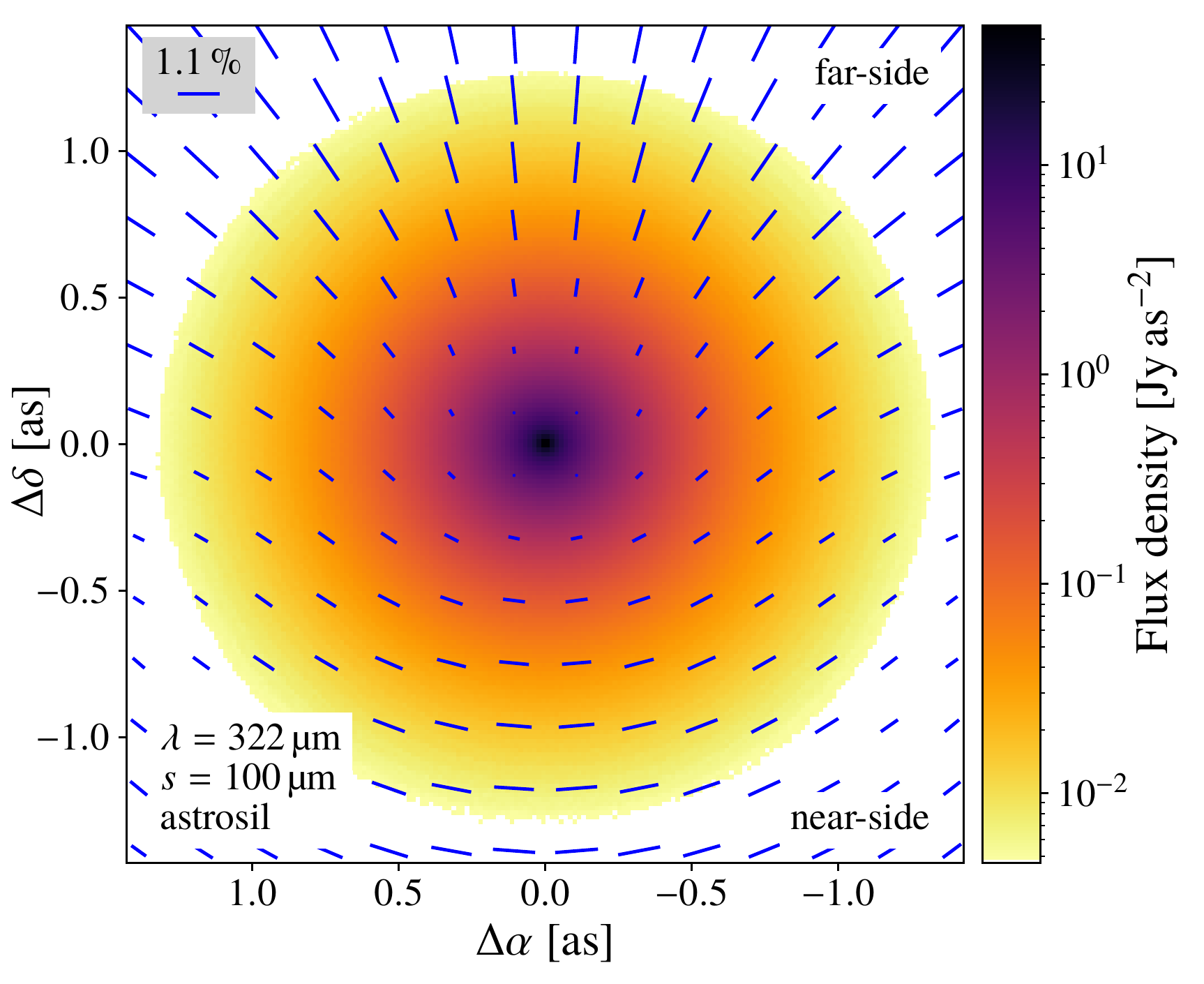}
            \includegraphics[width=0.495\linewidth]{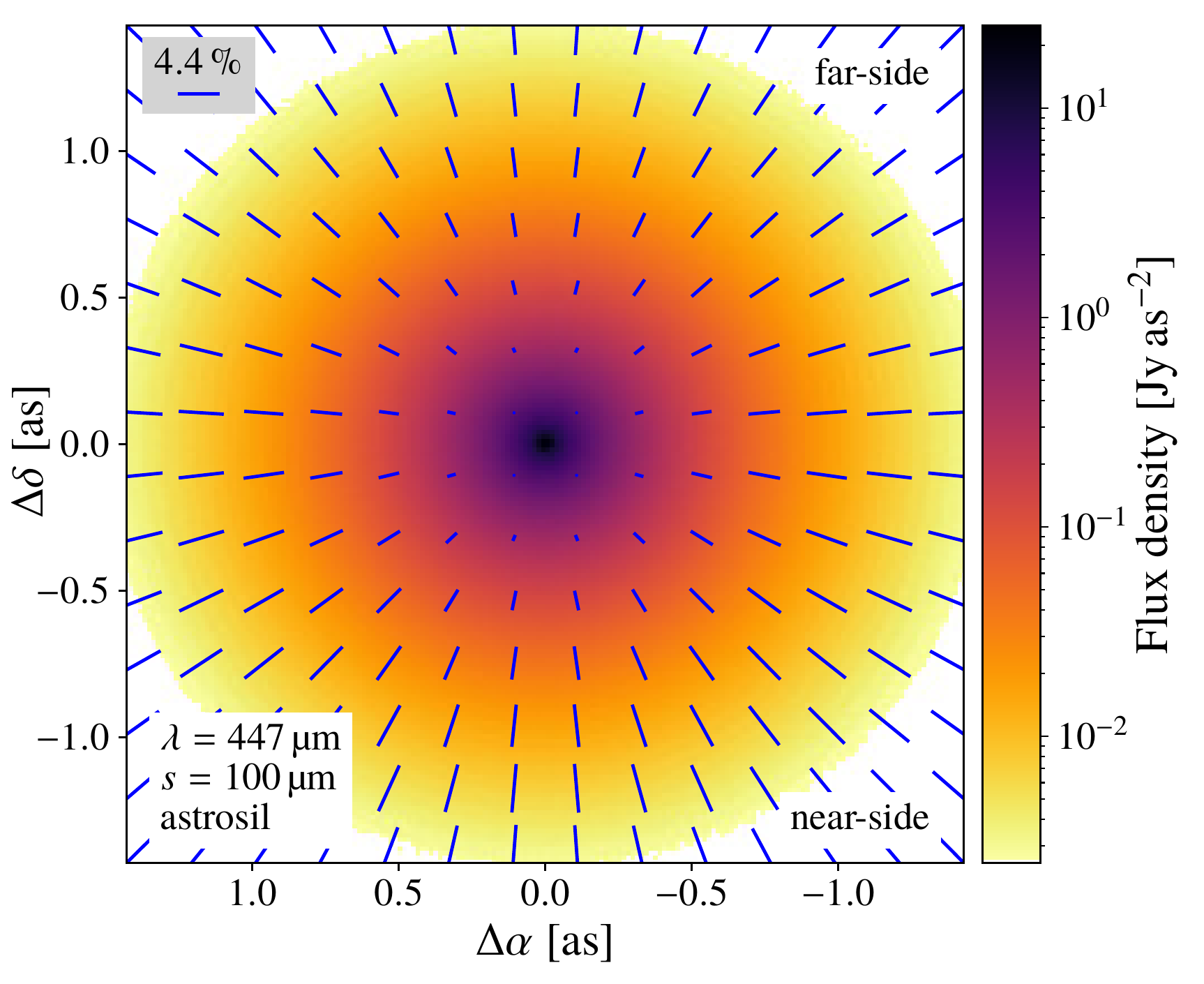}
            \caption{\textit{Top}: maximum degree of polarization $p$ of different wavelengths corresponding to the SOFIA/HAWC+ band \textit{D} ($\lambda=\SI{154}{\um}$), and the ALMA bands \textit{10} ($\lambda=\SI{300}{\um}$ and $\lambda=\SI{322}{\um}$) and \textit{9} ($\lambda=\SI{447}{\um}$). \textit{Middle and bottom:} corresponding intensity maps with superimposed polarization vectors. The inclination for all four sub-plots is $i=\SI{15}{\degree}$. The lengths of the polarization vectors are scaled to the maximum polarization degree present in each figure. Please note, that the resolution of SOFIA/HAWC+ is not sufficient to spatially resolve the polarization pattern of protoplanetary disks.}
            \label{fig:self_sca_incl}
        \end{figure}

        If the disk is slightly inclined, the dominating scattering angle will shift to a smaller (larger) value than \SI{90}{\degree} for the near-side (far-side) of the disk. Thus, asymmetric scattering will occur. This can be seen in \pref{fig:self_sca_incl} for the wavelengths $\lambda=\SI{154}{\um}$ (central wavelength of the SOFIA/HAWC+ \textit{D} band), $\lambda=\SIlist{300;322}{\um}$ (ALMA band \textit{10}), and $\lambda=\SI{447}{\um}$ (ALMA band \textit{9}). All disks are inclined by $i=\SI{15}{\degree}$ from face-on. In the SOFIA/HAWC+ \textit{D} band image (middle left), the near-side of the disk (lower half of the intensity maps) shows a larger polarization degree (as indicated by the length of the polarization vectors) than the far-side (upper half). This is due to the fact that $p = -\nicefrac{\mathcal{S}_{12}}{\mathcal{S}_{11}}$ is larger for scattering angles around $\theta = \SI{90}{\degree} - i = \SI{75}{\degree}$ than for scattering angles around $\theta = \SI{90}{\degree} + i = \SI{105}{\degree}$. If the disk were seen face-on, the polarization vectors would point radially outwards as $p$ is negative for $\theta = \SI{90}{\degree}$.
        
        The images for the wavelengths $\lambda=\SIlist{300;447}{\um}$ (ALMA bands \textit{10} and \textit{9}, respectively) show the polarization reversal and an additional asymmetry in the polarization degree. For $\lambda=\SI{300}{\um}$, the far-side shows a higher degree of polarization and for $\lambda=\SI{447}{\um}$, the near-side is slightly higher polarized.
        
\begin{figure}
            \includegraphics[width=0.495\hsize]{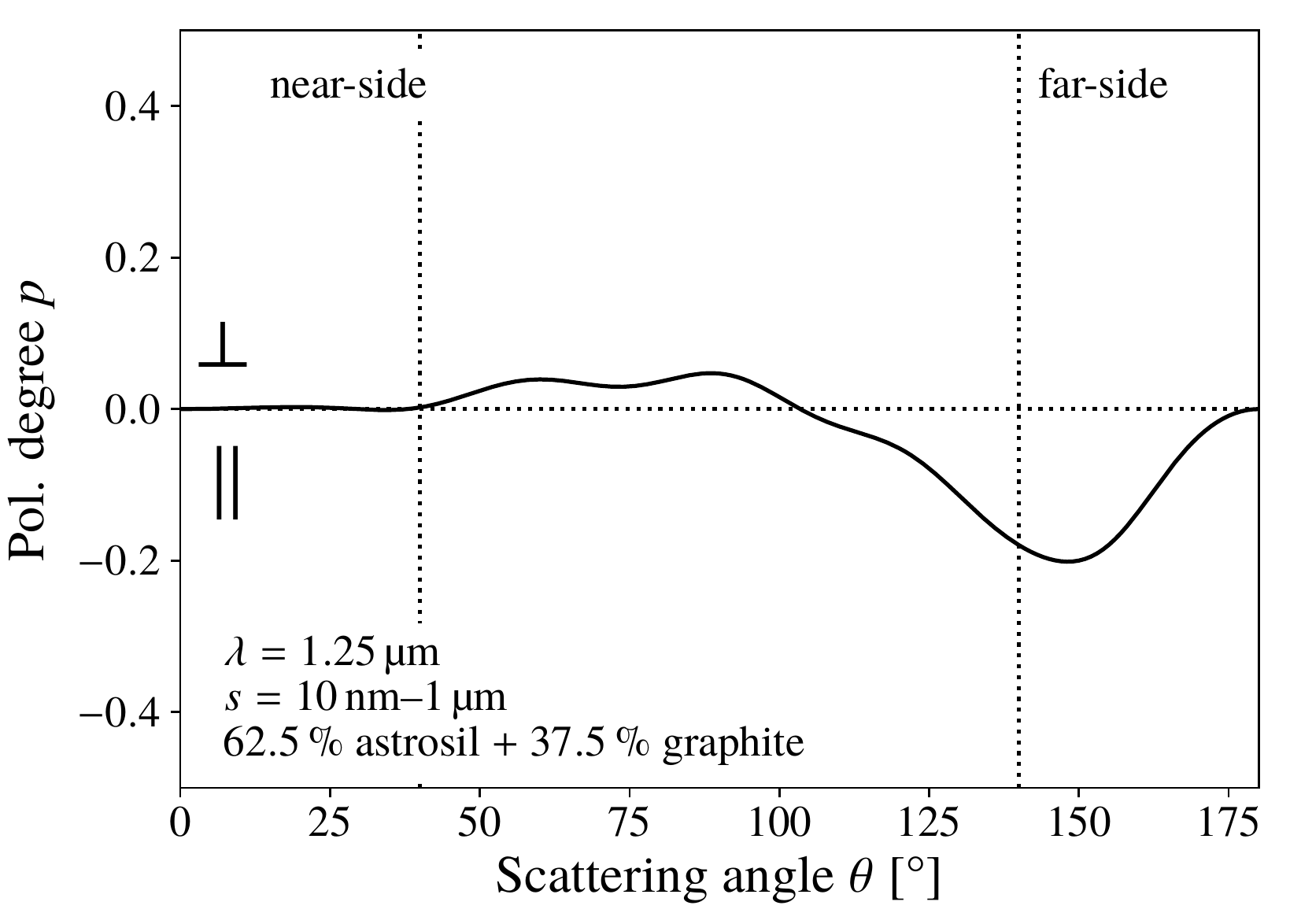}
            \includegraphics[width=0.495\hsize]{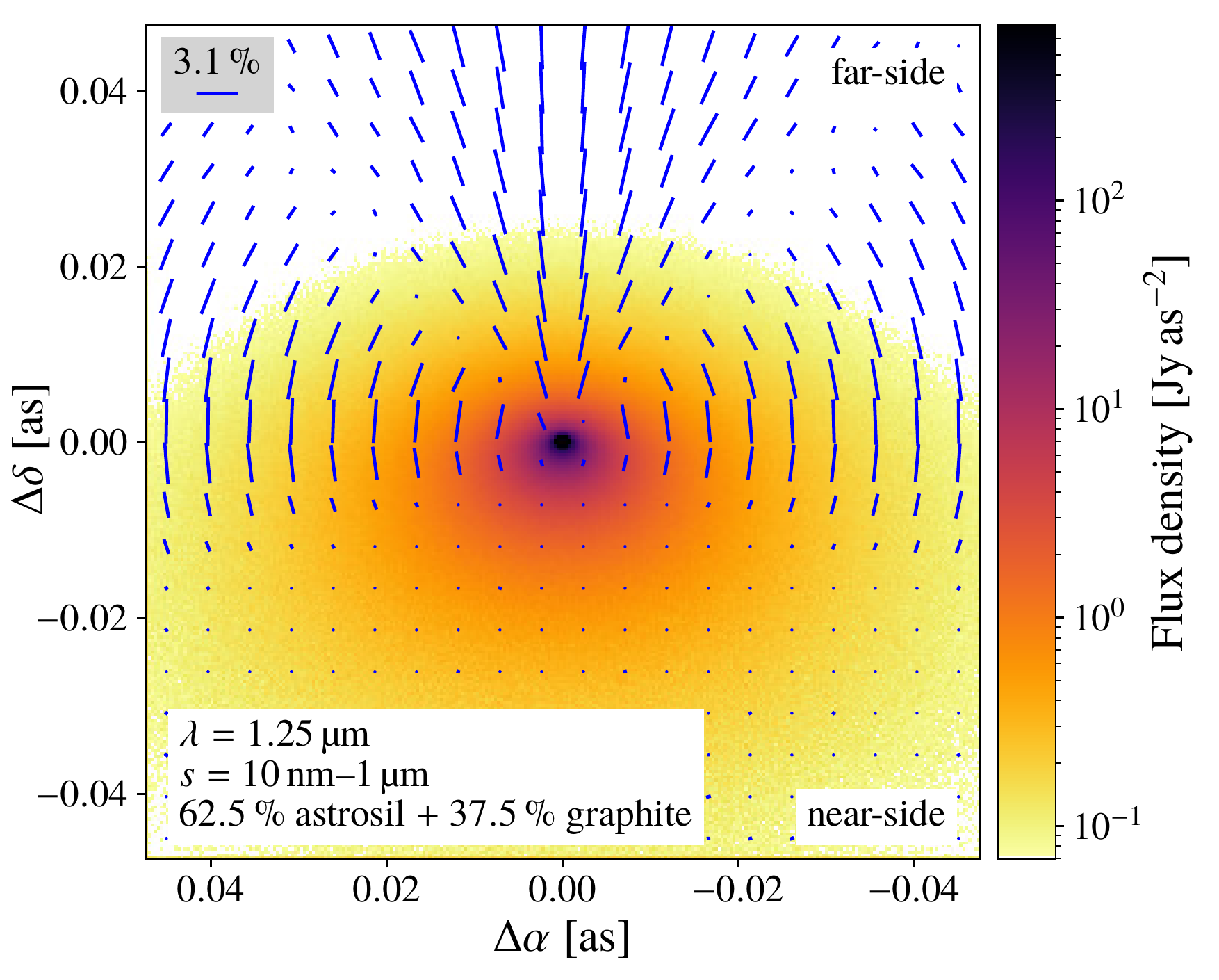}
            \caption{Polarization degree (\textit{left}) and intensity map for an inclined disk with an inclination of $i=\SI{50}{\degree}$ (\textit{right}) composed of a mixture of astronomical silicate and graphite of sizes between \SI{10}{\nano\m} and \SI{1}{\um} at a wavelength of $\lambda=\SI{1.25}{\um}$ (\textit{J} band). The dotted, vertical lines in the left panel represent the predominant scattering angles at the near- and far-side of the disk, i.e. $\theta=\SI{90}{\degree} \pm i$. The length of the polarization vectors are scaled to the maximum polarization degree present in the figure.}
            \label{fig:pol_frac_distro_mixture}
        \end{figure}

        A change of sign of the polarization fraction $p$ close to $\theta=\SI{90}{\degree}$ will result in different polarization patterns for the far- and near-side. This is clearly visible for the second wavelength of the ALMA band \textit{10} ($\lambda=\SI{322}{\um}$; see lower left panel of \pref{fig:self_sca_incl}). The polarization vectors at the far-side show the polarization reversal while the near-side shows a pattern of concentric rings because $p$ becomes negative for scattering angles larger than $\approx\SI{90}{\degree}$. The same effect but for a larger disk inclination of $i=\SI{50}{\degree}$ is shown in \pref{fig:pol_frac_distro_mixture}. This disk is composed of astronomical silicate and graphite with grain sizes from \SI{10}{\nano\m} to \SI{1}{\um} for an observing wavelength of $\lambda=\SI{1.25}{\um}$ (\textit{J} band).
        
        Thus, with precise measurements of the polarized light in the continuum, one is potentially able to stringently constrain the disk inclination even if it is as small as $\SI{15}{\degree}$.

\section{Conclusion}
\label{sec:conclusion}
    We investigated the influence of the anisotropic scattering function due to Mie scattering on the observable polarization pattern of protoplanetary disks resulting from scattered thermal dust emission in the mm/submm wavelength regime. If the signed polarization degree $p$ becomes negative, the resulting polarization pattern will undergo a \SI{90}{\degree} flip compared to the commonly assumed pattern due to scattering. This reversal is highly dependent on several parameters such as observing wavelength, grain size and composition, and disk inclination. Therefore, this effect, and the anisotropic character of the scattering function in general, may serve as a powerful tool to enable observers to set strict constraints on these parameters. This requires multi-wavelength polarization observations with high enough spatial and spectral resolution between roughly \SIrange[range-phrase={ and }]{100}{1000}{\um}. At the moment, only SOFIA is able to perform polarization measurements in the shorter wavelengths but with a restricted spatial resolution. For wavelengths larger than \SI{350}{\um}, ALMA is the perfect instrument to trace this polarization reversal.

\begin{acknowledgement}
    This research was funded through the DFG grant WO 857/18-1. RB thanks R.\,Brauer for providing general support in the application of \texttt{POLARIS}.
\end{acknowledgement}




\bibliographystyle{aa}
\bibliography{lit}

\begin{thebibliography}{35}
\expandafter\ifx\csname natexlab\endcsname\relax\def\natexlab#1{#1}\fi

\bibitem[{{Andrews} {et~al.}(2011){Andrews}, {Wilner}, {Espaillat}, {Hughes},
  {Dullemond}, {McClure}, {Qi}, \& {Brown}}]{andrews-et-al-2011}
{Andrews}, S.~M., {Wilner}, D.~J., {Espaillat}, C., {et~al.} 2011, \apj, 732,
  42

\bibitem[{{Bacciotti} {et~al.}(2018){Bacciotti}, {Girart}, {Padovani}, {Podio},
  {Paladino}, {Testi}, {Bianchi}, {Galli}, {Codella}, {Coffey}, {Favre}, \&
  {Fedele}}]{bacciotti-et-al-2018}
{Bacciotti}, F., {Girart}, J.~M., {Padovani}, M., {et~al.} 2018, \apjl, 865,
  L12

\bibitem[{{Blum} {et~al.}(2000){Blum}, {Wurm}, {Kempf}, {Poppe}, {Klahr},
  {Kozasa}, {Rott}, {Henning}, {Dorschner}, {Schr{\"a}pler}, {Keller},
  {Markiewicz}, {Mann}, {Gustafson}, {Giovane}, {Neuhaus}, {Fechtig},
  {Gr{\"u}n}, {Feuerbacher}, {Kochan}, {Ratke}, {El Goresy}, {Morfill},
  {Weidenschilling}, {Schwehm}, {Metzler}, \& {Ip}}]{blum-et-al-2000}
{Blum}, J., {Wurm}, G., {Kempf}, S., {et~al.} 2000, Physical Review Letters,
  85, 2426

\bibitem[{{Brunngr{\"a}ber} \& {Wolf}(2018)}]{brunngraeber-wolf-2018}
{Brunngr{\"a}ber}, R. \& {Wolf}, S. 2018, \aap, 611, A90

\bibitem[{{Brunngr{\"a}ber} {et~al.}(2016){Brunngr{\"a}ber}, {Wolf}, {Ratzka},
  \& {Ober}}]{brunngraeber-et-al-2016}
{Brunngr{\"a}ber}, R., {Wolf}, S., {Ratzka}, T., \& {Ober}, F. 2016, \aap, 585,
  A100

\bibitem[{{Casassus} {et~al.}(2018){Casassus}, {Avenhaus}, {P{\'e}rez},
  {Navarro}, {C{\'a}rcamo}, {Marino}, {Cieza}, {Quanz}, {Alarc{\'o}n}, {Zurlo},
  {Osses}, {Rannou}, {Rom{\'a}n}, \& {Barraza}}]{casassus-et-al-2018}
{Casassus}, S., {Avenhaus}, H., {P{\'e}rez}, S., {et~al.} 2018, \mnras, 477,
  5104

\bibitem[{{Daniel}(1980)}]{daniel-1980}
{Daniel}, J.-Y. 1980, \aap, 87, 204

\bibitem[{{Draine} \& {Flatau}(1994)}]{draine-flatau-1994}
{Draine}, B.~T. \& {Flatau}, P.~J. 1994, J. Opt. Soc. Am. A, 11, 1491

\bibitem[{{Draine} \& {Lee}(1984)}]{draine-lee-1984}
{Draine}, B.~T. \& {Lee}, H.~M. 1984, \apj, 285, 89

\bibitem[{{Draine} \& {Malhotra}(1993)}]{draine-malhotra-1993}
{Draine}, B.~T. \& {Malhotra}, S. 1993, \apj, 414, 632

\bibitem[{{Fischer} {et~al.}(1994){Fischer}, {Henning}, \&
  {Yorke}}]{fischer-et-al-1994}
{Fischer}, O., {Henning}, T., \& {Yorke}, H.~W. 1994, \aap, 284, 187

\bibitem[{{Glauser} {et~al.}(2008){Glauser}, {M{\'e}nard}, {Pinte},
  {Duch{\^e}ne}, {G{\"u}del}, {Monin}, \& {Padgett}}]{glauser-et-al-2008}
{Glauser}, A.~M., {M{\'e}nard}, F., {Pinte}, C., {et~al.} 2008, \aap, 485, 531

\bibitem[{{Hartmann} {et~al.}(1998){Hartmann}, {Calvet}, {Gullbring}, \&
  {D'Alessio}}]{hartmann-et-al-1998}
{Hartmann}, L., {Calvet}, N., {Gullbring}, E., \& {D'Alessio}, P. 1998, \apj,
  495, 385

\bibitem[{{Hull} {et~al.}(2018){Hull}, {Yang}, {Li}, {Kataoka}, {Stephens},
  {Andrews}, {Bai}, {Cleeves}, {Hughes}, {Looney}, {P{\'e}rez}, \&
  {Wilner}}]{hull-et-al-2018}
{Hull}, C.~L.~H., {Yang}, H., {Li}, Z.-Y., {et~al.} 2018, \apj, 860, 82

\bibitem[{{Kataoka} {et~al.}(2016){Kataoka}, {Muto}, {Momose}, {Tsukagoshi}, \&
  {Dullemond}}]{kataoka-et-al-2016}
{Kataoka}, A., {Muto}, T., {Momose}, M., {Tsukagoshi}, T., \& {Dullemond},
  C.~P. 2016, \apj, 820, 54

\bibitem[{{Kataoka} {et~al.}(2015){Kataoka}, {Muto}, {Momose}, {Tsukagoshi},
  {Fukagawa}, {Shibai}, {Hanawa}, {Murakawa}, \&
  {Dullemond}}]{kataoka-et-al-2015}
{Kataoka}, A., {Muto}, T., {Momose}, M., {et~al.} 2015, \apj, 809, 78

\bibitem[{{Keppler} {et~al.}(2018){Keppler}, {Benisty}, {M{\"u}ller},
  {Henning}, {van Boekel}, {Cantalloube}, {Ginski}, {van Holstein}, {Maire},
  {Pohl}, {Samland}, {Avenhaus}, {Baudino}, {Boccaletti}, {de Boer},
  {Bonnefoy}, {Chauvin}, {Desidera}, {Langlois}, {Lazzoni}, {Marleau},
  {Mordasini}, {Pawellek}, {Stolker}, {Vigan}, {Zurlo}, {Birnstiel},
  {Brandner}, {Feldt}, {Flock}, {Girard}, {Gratton}, {Hagelberg}, {Isella},
  {Janson}, {Juhasz}, {Kemmer}, {Kral}, {Lagrange}, {Launhardt}, {Matter},
  {M{\'e}nard}, {Milli}, {Molli{\`e}re}, {Olofsson}, {P{\'e}rez}, {Pinilla},
  {Pinte}, {Quanz}, {Schmidt}, {Udry}, {Wahhaj}, {Williams}, {Buenzli},
  {Cudel}, {Dominik}, {Galicher}, {Kasper}, {Lannier}, {Mesa}, {Mouillet},
  {Peretti}, {Perrot}, {Salter}, {Sissa}, {Wildi}, {Abe}, {Antichi},
  {Augereau}, {Baruffolo}, {Baudoz}, {Bazzon}, {Beuzit}, {Blanchard}, {Brems},
  {Buey}, {De Caprio}, {Carbillet}, {Carle}, {Cascone}, {Cheetham}, {Claudi},
  {Costille}, {Delboulb{\'e}}, {Dohlen}, {Fantinel}, {Feautrier}, {Fusco},
  {Giro}, {Gluck}, {Gry}, {Hubin}, {Hugot}, {Jaquet}, {Le Mignant}, {Llored},
  {Madec}, {Magnard}, {Martinez}, {Maurel}, {Meyer}, {M{\"o}ller-Nilsson},
  {Moulin}, {Mugnier}, {Orign{\'e}}, {Pavlov}, {Perret}, {Petit}, {Pragt},
  {Puget}, {Rabou}, {Ramos}, {Rigal}, {Rochat}, {Roelfsema}, {Rousset}, {Roux},
  {Salasnich}, {Sauvage}, {Sevin}, {Soenke}, {Stadler}, {Suarez}, {Turatto}, \&
  {Weber}}]{keppler-et-al-2018}
{Keppler}, M., {Benisty}, M., {M{\"u}ller}, A., {et~al.} 2018, \aap, 617, A44

\bibitem[{{Kirchschlager} \& {Wolf}(2013)}]{kirchschlager-wolf-2013}
{Kirchschlager}, F. \& {Wolf}, S. 2013, \aap, 552, A54

\bibitem[{{Kirchschlager} \& {Wolf}(2014)}]{kirchschlager-wolf-2014}
{Kirchschlager}, F. \& {Wolf}, S. 2014, \aap, 568, A103

\bibitem[{{Laor} \& {Draine}(1993)}]{laor-draine-1993}
{Laor}, A. \& {Draine}, B.~T. 1993, \apj, 402, 441

\bibitem[{{Lee} {et~al.}(2018){Lee}, {Li}, {Ching}, {Lai}, \&
  {Yang}}]{lee-et-al-2018}
{Lee}, C.-F., {Li}, Z.-Y., {Ching}, T.-C., {Lai}, S.-P., \& {Yang}, H. 2018,
  \apj, 854, 56

\bibitem[{{Lynden-Bell} \& {Pringle}(1974)}]{lynden-bell-pringle-1974}
{Lynden-Bell}, D. \& {Pringle}, J.~E. 1974, \mnras, 168, 603

\bibitem[{{Mathis} {et~al.}(1977){Mathis}, {Rumpl}, \&
  {Nordsieck}}]{mathis-et-al-1977}
{Mathis}, J.~S., {Rumpl}, W., \& {Nordsieck}, K.~H. 1977, \apj, 217, 425

\bibitem[{{Ohashi} {et~al.}(2018){Ohashi}, {Kataoka}, {Nagai}, {Momose},
  {Muto}, {Hanawa}, {Fukagawa}, {Tsukagoshi}, {Murakawa}, \&
  {Shibai}}]{ohashi-et-al-2018}
{Ohashi}, S., {Kataoka}, A., {Nagai}, H., {et~al.} 2018, \apj, 864, 81

\bibitem[{{Ormel} {et~al.}(2008){Ormel}, {Cuzzi}, \&
  {Tielens}}]{ormel-et-al-2008}
{Ormel}, C.~W., {Cuzzi}, J.~N., \& {Tielens}, A.~G.~G.~M. 2008, \apj, 679, 1588

\bibitem[{{Reissl} {et~al.}(2016){Reissl}, {Wolf}, \&
  {Brauer}}]{reissl-et-al-2016}
{Reissl}, S., {Wolf}, S., \& {Brauer}, R. 2016, \aap, 593, A87

\bibitem[{{Stephens} {et~al.}(2014){Stephens}, {Looney}, {Kwon},
  {Fern{\'a}ndez-L{\'o}pez}, {Hughes}, {Mundy}, {Crutcher}, {Li}, \&
  {Rao}}]{stephens-et-al-2014}
{Stephens}, I.~W., {Looney}, L.~W., {Kwon}, W., {et~al.} 2014, \nat, 514, 597

\bibitem[{{Varga} {et~al.}(2018){Varga}, {{\'A}brah{\'a}m}, {Chen}, {Ratzka},
  {Gab{\'a}nyi}, {K{\'o}sp{\'a}l}, {Matter}, {van Boekel}, {Henning}, {Jaffe},
  {Juh{\'a}sz}, {Lopez}, {Menu}, {Mo{\'o}r}, {Mosoni}, \&
  {Sipos}}]{varga-et-al-2018}
{Varga}, J., {{\'A}brah{\'a}m}, P., {Chen}, L., {et~al.} 2018, \aap, 617, A83

\bibitem[{{Warren} \& {Brandt}(2008)}]{warren-brandt-2008}
{Warren}, S.~G. \& {Brandt}, R.~E. 2008, Journal of Geophysical Research:
  Atmospheres, 113
  [\eprint{https://agupubs.onlinelibrary.wiley.com/doi/pdf/10.1029/2007JD009744}]

\bibitem[{{Weingartner} \& {Draine}(2001)}]{weingartner-draine-2001}
{Weingartner}, J.~C. \& {Draine}, B.~T. 2001, \apj, 548, 296

\bibitem[{{Williams} \& {Cieza}(2011)}]{williams-cieza-2011}
{Williams}, J.~P. \& {Cieza}, L.~A. 2011, Annual Review of Astronomy and
  Astrophysics, 49, 67

\bibitem[{{Wolf} {et~al.}(2008){Wolf}, {Schegerer}, {Beuther}, {Padgett}, \&
  {Stapelfeldt}}]{wolf-et-al-2008}
{Wolf}, S., {Schegerer}, A., {Beuther}, H., {Padgett}, D.~L., \& {Stapelfeldt},
  K.~R. 2008, \apjl, 674, L101

\bibitem[{{Yang} {et~al.}(2016{\natexlab{a}}){Yang}, {Li}, {Looney}, \&
  {Stephens}}]{yang-et-al-2016a}
{Yang}, H., {Li}, Z.-Y., {Looney}, L., \& {Stephens}, I. 2016{\natexlab{a}},
  \mnras, 456, 2794

\bibitem[{{Yang} {et~al.}(2016{\natexlab{b}}){Yang}, {Li}, {Looney}, {Cox},
  {Tobin}, {Stephens}, {Segura-Cox}, \& {Harris}}]{yang-et-al-2016b}
{Yang}, H., {Li}, Z.-Y., {Looney}, L.~W., {et~al.} 2016{\natexlab{b}}, \mnras,
  460, 4109

\bibitem[{{Yang} {et~al.}(2017){Yang}, {Li}, {Looney}, {Girart}, \&
  {Stephens}}]{yang-et-al-2017}
{Yang}, H., {Li}, Z.-Y., {Looney}, L.~W., {Girart}, J.~M., \& {Stephens}, I.~W.
  2017, \mnras, 472, 373

\end{thebibliography}



\begin{appendix}
    \section{Disk set-up}
    \label{sec:disk_setup}
        The density distribution of the dust is based on the studies by \citet{lynden-bell-pringle-1974}, \citet{hartmann-et-al-1998}, and \citet{andrews-et-al-2011} and given by
        \begin{equation}
        \label{eq:dens_distro}
            \varrho(r) = \frac{\Sigma(r)}{\sqrt{2\pi}\,h(r)}\!\cdot\exp{\left[-\frac{1}{2}\left(\frac{z}{h(r)}\right)^2\right]}
        \end{equation}
        with the vertical integrated surface density
        \begin{equation*}
            \Sigma(r) = \sqrt{2\pi}\,\rho_0h_{\text{ref}}\cdot\left(\frac{r}{R_{\text{ref}}}\right)^{-\gamma}\cdot\exp{\left[-\left(\frac{r}{R_{\text{ref}}}\right)^{2-\gamma}\right]}
        \end{equation*}
        and the scale height
        \begin{equation*}
            h(r) = h_{\text{ref}}\left(\frac{r}{R_{\text{ref}}}\right)^{\beta}\ .
        \end{equation*}
        In \pref{eq:dens_distro}, $r$ and $z$ are the usual cylindrical coordinates, and $\rho_0$ is the number density at $r=R_{\text{ref}}$ and $z=\SI{0}{\au}$ and is scaled to the given total dust mass. The disk parameter values used in this study are compiled in \pref{tab:param_space}.
        
        \begin{table}
            \caption{Disk parameter values.}
            \label{tab:param_space}
            \centering
            \begin{tabular}{l l l}
                \hline\hline
                \rule{0pt}{2.5ex}Parameter    & Variable                       & Values    \\[1mm]
                \hline
                \rule{0pt}{2.5ex}Inner radius & $R_{\text{in}}$ [\si{\au}]     & \num{0.1} \\
                Outer radius                  & $R_{\text{out}}$ [\si{\au}]    & \num{300} \\
                Reference radius              & $R_{\text{ref}}$ [\si{\au}]    & \num{100} \\
                Reference scale height        & $h_{\text{ref}}$ [\si{\au}]    & \num{10}  \\
                Density profile               & $\gamma$                       & \num{0.8} \\
                Flaring parameter             & $\beta$                        & \num{1.1} \\
                Dust mass                     & $M_{\text{dust}}$ [\si{\msun}] & \num{e-4} \\
                Distance                      & $d$ [\si{\pc}]                 & \num{140} \\
                \hline
            \end{tabular}
        \end{table}
\end{appendix}

\end{document}